\documentstyle[pra,aps,epsfig]{revtex}

\input epsf

\begin{document}

%%\twocolumn[\hsize\textwidth\columnwidth\hsize\csname
%%@twocolumnfalse\endcsname
%%]

\title{Quest for Fast  Partial Search Algorithm}

\author{ Vladimir\ E.\ Korepin$^1$ and Jinfeng Liao$^2$}
\address{ $^1$C.N. Yang Institute for Theoretical Physics, State
 University of New York at Stony Brook, Stony Brook, NY 11794-3840  \\
$^2$Department of Physics and Astronomy, State University of New
York at Stony Brook, Stony Brook, NY 11794-3800
 }

\maketitle

\begin{abstract}
A quantum algorithm can find a target item in a database faster
than a classical algorithm. One can trade accuracy for speed and
find a part of the database (a block) containing the target item
even  faster, this is  partial search. We consider different
partial search algorithms and suggest the optimal one. Efficiency
of an algorithm is measured by number of queries to the oracle.
\end{abstract}\vspace{0.1in}

\section{Introduction}

Database search has many applications and is used widely. Grover
discovered a quantum algorithm that searches faster than a
classical algorithm \cite{Grover}.  It consists of repetition of
the Grover iteration $\hat{G}_1$, which operates on the
computational quantum states.
% Recently nuclear magnetic resonance (NMR) was used to implement it \cite{jx}
The number of  repetitions is:
\begin{equation}
j_{\mbox{full}}= \frac{\pi}{4} \sqrt{N} \label{full}
\end{equation}
for a database with large number of entries $N$. After
$j_{\mbox{full}} $ the algorithm finds the target item. For more
details, see also \cite{Bennett,Boyer,Brass}. Below we shall call
$\hat{G}_1$ a global iteration.

Sometimes  it is sufficient to find an approximate location of
the target item. A partial search considers the following problem:
a  database is
 separated into
%\begin{equation}
$K$ { blocks},
%\end{equation}
of  a  size
%\begin{equation}
$b={N}/{K}$.
%\end{equation}
%\footnote{Here we only consider a case when both $N$ and $K$ are powers of $2$.}
We  want to find a block with the target item, not the target item
itself. Such partial search  was first introduced by Mark Heiligman in
\cite{hei}, as a part of algorithm for list matching.
 We can think of partial search  in following terms: an exact address
of the target item is given by a sequence of $n$ bites , but we
want to find only first $k$ bites ($k<n$). 
%It was first introduced by Mark Heiligman in \cite{hei},
%as a part of algorithm for list matching.
Fast  quantum algorithm for a partial search was found by Grover and
Radhakrishnan in \cite{jaik}. They showed that  classical partial
search takes  $\sim (N-b)$ queries, but
% but  quantum is $\sim \sqrt{b} $ faster then full search.
quantum  algorithm takes only $\sim ( \sqrt{N}-
\mbox{coeff}\sqrt{b}) $ queries.
% and simplified in  \cite{kg}.
% and simplified in \cite{kg}.
 It uses several global iterations $\hat{G}^{j_1}_1$ and
then several local
 iteration $\hat{G}^{j_2}_2$, see (\ref{liter}). Local searches are Grover iterations [searches] in each individual block made in
each block separately in parallel. Grover-Radhakrishnan algorithm
was improved and simplified in  \cite{kg}. The number of queries to the
oracle in this   algorithm was minimized
 by  in \cite{kor}, the $ \mbox{coeff}$ was maximized.
Below we shall explain minimized version of Grover-Radhakrishnan
algorithm.
 We shall call it GRK algorithm.
In this paper we consider three other versions of partial search
algorithm. They use different sequences of global and local
searches: local-global, global-local-global and
local-global-local. We prove that GRK version  still uses  minimal
number of queries to the oracle. We conjecture that{\it  GRK
  algorithm is optimal among all partial search algorithms, which consist of
 arbitrary sequence of local and global searches}.

The plan of the paper is as follows.  In the next section we
remind the Grover algorithm. After this  we
 formulate minimized version of Grover-Radhakrishnan algorithm
[GRK algorithm]. In the rest of the paper we consider other
partial search algorithms. We arrive at the conclusion that GRK
uses minimal number of queries comparing to other algorithms.
%The algorithm  consists of three steps, described below.
%For  many large blocks   the   algorithm  was  improved   in  \cite{kg}.
%In this paper  we  optimize the  algorithm.
%    partial search algorithm  and compare it with
% other approaches for different numbers of blocks, denoted $K$.
%We also improve a lower bound for number of queries on an oracle, comparing
%to \cite{jaik}.
%We modify the partial search algorithm in such a way that the final state of
%the target block can be represented as a result of several applications of
%Grover iterations (\ref{iter}) to the standard initial state (\ref{ave}); see
%eq. (\ref{usk}).  If a somebody will decide to make another quantum search [to
%use the target block as a new database] he/she can do it faster than
%(\ref{full}), as will be shown in eq. (\ref{luch}).

\section{Partial search}
\subsection{Global Iterations}
First let us remind the full Grover search. We shall consider a
database with one target item. The aim of the Grover algorithm is
to identify a target state $|t\rangle$ among an unordered set of
$N$ states. This is achieved by repeating global iteration which
is defined in terms of two operators. The first changes the sign
of the target state $|t\rangle$ only:
\begin{equation}
\hat{I_t}=\hat{I}-2|t\rangle \langle t|, \qquad \langle
t|t\rangle=1, \label{target}
\end{equation}
where $\hat{I}$ is the identity operator. The second operator,
\begin{equation}
\hat{I}_{s_1}=\hat{I}-2|s_1\rangle \langle s_1|, \label{average}
\end{equation}
changes the sign of the uniform superposition of all basis states
$|s_1\rangle$,
\begin{equation}
|s_1\rangle = \frac{1}{\sqrt{N}}\sum_{x=0}^{N-1}|x\rangle , \qquad
\langle s_1|s_1\rangle =1 . \label{ave}
\end{equation}
%The operator $I_{s_1}$ reflects amplitudes $ a_x$ about the average:
%\begin{eqnarray}
%-I_{s_1}\sum a_x|x\rangle
%=\sum_{x=0}^{N-1}(2\bar{a}-a_x)|x\rangle,
%\mbox{\u{a}}_x =
% \quad \bar{a}=\sum_{x=0}^{N-1}
%\frac{a_x}{N} \label{reflect}
%\end{eqnarray}
The {\bf global iteration} is defined as a unitary operator
\begin{equation}
\hat{G_1}=-\hat{I}_{s_1}\hat{I}_t . \label{iter}
\end{equation}
%and will be called a global iteration.
We shall use  eigenvectors  of $\hat{G_1}$:
%\begin{equation}
%Q|\psi^{\pm}_1\rangle =\lambda^{\pm}_1 |\psi^{\pm}_1\rangle
%\end{equation}
\begin{eqnarray}
\hat{G_1}|\psi^{\pm}_1\rangle  = \lambda^{\pm}_1
|\psi^{\pm}_1\rangle , \qquad \lambda^{\pm}_1 =\exp[{\pm 2i
\theta_1}] , \qquad
 |\psi^{\pm}_1\rangle  = \frac{1}{\sqrt{2}}|t\rangle \pm \frac{i}
{\sqrt{2}}\left(
\sum^{N-1}_{\stackrel{\mbox{\small{x=0}}}{\mbox{\small{x  $\neq$
t}}}} \frac{|x\rangle}{\sqrt{(N-1)}} \right). \label{value}
% \lambda^{\pm}_1& =&\exp[{\pm 2i \theta_1}] . \label{value}
\end{eqnarray}
They were found in \cite{Brass}, where the angle $\theta_1$ is
defined by
\begin{equation}
\sin ^2\theta_1 =\frac{1}{N}. \label{ang1}
\end{equation}

%At the  Step 1  we  apply  $j_1$ successive Grover iterations
%to the initial state (\ref{ave})
%After $j$ iterations, the result is \cite{Brass}
%\begin{eqnarray}
%G_1^{j_1} |s_1\rangle  = \sin \left( (2j_1+1)\theta_1 \right) & |t\rangle& +
%\nonumber \\
% \frac{\cos \left( (2j_1+1)\theta_1 \right)}{\sqrt{N-1}}
%\sum^{N-1}_{\stackrel{\mbox{\small{x=0}}}{\mbox{\small{x  $\neq$  t}}}}&
%|x\rangle& ,
%\label{first}
%\end{eqnarray}

\subsection{GRK Algorithm  for Partial Search}

The first version of partial search was found in \cite{jaik}. The
algorithm uses $j_1$ global iteration and $j_2$ local iterations.
{\bf LOCAL ITERATIONS}  are Grover iterations for each block:
\begin{equation}
\hat{G}_2=-\hat{I}_{s_2}\hat{I}_t .\label{liter}
\end{equation}
$\hat{I}_t $ is given by (\ref{target}), but $\hat{I}_{s_2}$ is
different. In one block it acts as:
\begin{equation}\label{is2}
\hat{I}_{s_2}{\big |}_{block}=\hat{I}{\big |}_{block}-2|s_2\rangle
\langle s_2|, \qquad |s_2\rangle =
\frac{1}{\sqrt{b}}\sum_{\mbox{\scriptsize{one block}}}|x\rangle
\label{nblock}\end{equation}
%where we sum up only within one block.
In the whole database  $\hat{I}_{s_2}$  is the direct sum of
(\ref{is2}) with respect to all blocks. Both relevant
eigenvectors of $\hat{G_2}$ were found by in \cite{Brass}:
\begin{eqnarray}
\hat{G}_2|\psi^{\pm}_2\rangle =\lambda^{\pm}_2
|\psi^{\pm}_2\rangle, \qquad
 \lambda^{\pm}_2  = \exp[{\pm 2i \theta_2}] , \label{vector2}  \qquad
|\psi^{\pm}_2\rangle   =\frac{1}{\sqrt{2}}|t\rangle \pm
\frac{i}{\sqrt{2}} |\mbox{ntt}\rangle
% \lambda^{\pm}_2 & =& \exp[{\pm 2i \theta_2}] .
\end{eqnarray}

Here the $|\mbox{ntt}\rangle $ is a normalized sum of all
non-target items in the target block:
\begin{equation}
|\mbox{ntt}\rangle = \frac{1}{\sqrt{b-1}} \sum_{\stackrel{x \neq
t}{\mbox{\tiny{target block}}}} |x\rangle,\qquad \langle
\mbox{ntt} |\mbox{ntt}\rangle =1 .\label{vntt}
\end{equation}
We shall need  an angle $\theta_2$ given by
\begin{equation}
\sin^2 \theta_2 =\frac{K}{N}=\frac{1}{b}. \label{ang2}
\end{equation}

The partial search  algorithm of \cite{jaik} creates a vector
\begin{equation}
|d\rangle =\hat{G}_1 \hat{G}^{j_1}_2 \hat{G}^{j_0}_1|s_1\rangle
.\label{gl}
\end{equation}
see \footnote{ We  use a  modification of  \cite{jaik}, suggested
in \cite{kor}}. In the state $|d\rangle $ the amplitudes of all
items in non-target blocks are zero. Notice that  this algorithm
uses  global-local sequence of searches. We consider large blocks
$b=N/K\rightarrow \infty$. The number of blocks $K$ is an
important parameter. We shall replace it with
\begin{equation}
\sin \gamma =\frac{1}{\sqrt{K}}, \qquad 0\le \gamma \le
\frac{\pi}{4} \label{gam}
\end{equation}
The optimal version of this algorithm was find in \cite{kor}. It
can be described by the following equations:
\begin{eqnarray}
\cos (2j_1 \theta_2)=\frac{\sin \gamma \cos 2\gamma }{\cos \gamma
\sin 2\gamma}, \qquad
 \tan (2j_0 \theta_1)&=& \frac{\cos 2\gamma }{(\sin \gamma) \sqrt{3-4
(\sin \gamma)^2} } \label{korepin}
\end{eqnarray}
Partial search is faster then full search (\ref{full}) by $\sim
\sqrt{b}$, the coefficient in front of $\sqrt{b}$ is explicitly
calculated in \cite{kor} and studied as a function of number of
blocks.

\subsection{Notation and Setup for General Partial Search Algorithm }

Let us introduce a unite vector:
\begin{eqnarray}
%|t\rangle &,& \qquad |ntt\rangle   \\
|u\rangle = \frac{1}{\sqrt{b(K-1)}} \sum_{\stackrel{\mbox{all
items in all}}{\mbox{{non-target blocks}}}} |x\rangle , \qquad
\langle u|u\rangle =1 .\label{ntb}
\end{eqnarray}
We shall use a three dimensional space. The orthonormal basis is
formed by the target item $|t\rangle$, sum of all non-target items
in the target block $|ntt\rangle$, defined in (\ref{vntt}) and
$|u\rangle $. All the state vectors involved in present quantum
search problem can be written in this basis as
\begin{equation} |V>=(a,b,c)^T \end{equation}
meaning
\begin{equation}\label{vectorform}|V>=a|t>+b|ntt>+c|u>\end{equation} For
example, the global uniform state which is used as initial state
of searching is given by
\begin{equation}
|s_1\rangle =( \sin \gamma \sin \theta_2, \sin \gamma \cos
\theta_2 , \cos \gamma )^T \label{initial}
\end{equation}
and the local uniform state is
\begin{equation}
|s_2\rangle =( \sin \theta_2, \cos \theta_2 , 0 )^T \label{local
initial}
\end{equation}

The algorithms which we consider in this paper can be represented
as matrices in this linear space. For example $j_2$ repetitions of
the local iteration (\ref{liter}) is:
\begin{equation}
\hat{G}^{j_2}_2=\left( \begin{array}{clcr}
\cos (2j_2\theta_2) & \sin (2j_2\theta_2) &0 \\
-\sin (2j_2\theta_2) & \cos (2j_2\theta_2) &0 \\
0 & 0& 1
\end{array}
\right) \label{loc}
\end{equation}
The ordering of eigenvectors is  $|t\rangle$, $|ntt\rangle $  and
$|u\rangle $. The matrix  has three eigenvectors:
\begin{eqnarray}
\hat{G}^{j_2}_2|v_2^{\pm}\rangle =\exp (\pm 2i\theta_2
j_2)|v_2^{\pm}\rangle ,\quad \hat{G}^{j_2}_2|v_2^{0}\rangle
=|v_2^{0}\rangle
\end{eqnarray}
The eigenvectors can be represented as:
\begin{equation}
|v_2^{\pm}\rangle=\frac{1}{\sqrt{2}}\left(\begin{array}{c}
1\\
\pm i \\
0
\end{array} \right), \qquad |v_2^{0}\rangle= \left(\begin{array}{c}
0\\
0 \\
1
\end{array} \right).
\end{equation}

Now let us turn our attention to global iterations (\ref{iter}),
$j_1$ repetitions of the global iterations can be represented as
\begin{equation}
\hat{G}^{j_1}_1= \left( \begin{array}{clcr} \cos (2j_1\theta_1), &
\sin (2j_1\theta_1) \sin \gamma, &
\sin (2j_1\theta_1) \cos \gamma \\
-\sin (2j_1\theta_1)\sin \gamma,\quad  & (-1)^{j_1}\cos^2 \gamma +\cos (2j_1\theta_1) \sin^2 \gamma,\quad &\sin \gamma \cos \gamma \left( (-1)^{j_1+1}+\cos (2j_1\theta_1)\right) \\
- \sin (2j_1\theta_1) \cos \gamma, \quad &\sin \gamma \cos \gamma
\left( (-1)^{j_1+1}+\cos (2j_1\theta_1)\right),  &
(-1)^{j_1}\sin^2 \gamma +\cos (2j_1\theta_1) \cos^2 \gamma
\end{array}
\right)
\end{equation}
This is a simplified asymptotic expression valid in the limit of
large blocks $b\rightarrow \infty$. We used (\ref{gam}). The
matrix  has three eigenvectors:
\begin{eqnarray}
\hat{G}^{j_1}_1|v_1^{\pm}\rangle =\exp (\pm 2i\theta_1
j_1)|v_1^{\pm}\rangle ,\quad \hat{G}^{j_1}_1|v_1^{0}\rangle
=(-1)^{j_1}|v_1^{0}\rangle
\end{eqnarray}
The eigenvectors can be represented as:
\begin{equation}
|v_1^{\pm}\rangle=\frac{1}{\sqrt{2}}\left(\begin{array}{c}
1\\
\pm i\sin \gamma \\
\pm i\cos \gamma
\end{array} \right), \qquad |v_1^{0}\rangle= \left(\begin{array}{c}
0\\
\cos \gamma \\
-\sin \gamma
\end{array} \right).
\end{equation}

\subsection{General Problem of Partial Search}
The ultimate goal of partial search is to start with the uniform
state $|s_1>$ and locate the target block, and obviously GRK is
not the only means to achieve the goal. Based on the two types of
queries, global and local iterations, we naturally generalize GRK
into a wide set of partial search algorithms by alternate use of
the two iterations:
\begin{equation}
\hat{G}(j_k,j_{k-1},\cdot\cdot\cdot,j_2,j_1,j_0)=\hat{G_1}^{j_k}
\hat{G_2}^{j_{k-1}} \hat{G_1}^{j_{k-2}} \hat{G_2}^{j_{k-3}} \cdot
\cdot \cdot \hat{G_1}^{j_2} \hat{G_2}^{j_1} \hat{G_1}^{j_0}
\end{equation}
which fulfills the following condition
\begin{equation} \label{constraint}
<u|\, \hat{G}(j_k,j_{k-1},\cdot\cdot\cdot,j_2,j_1,j_0) \, |s_1>=0
\end{equation}
This means that amplitude of each item in non-target block is
zero. Here all $j_i$ are non-negative integers. The total number
of queries for these algorithms is given by
$S=\sum_{i=0}^{_k}j_i$, and we should try to minimize $S$ to find
the optimal one.

Now let's consider various sequences. A few discussions can be
made here:\\
1) The last queries in the sequence should be the global
iterations. Note that local iteration (\ref{loc}) doesn't do
anything on $|u>$ but only rotates state component inside the
target block, so if an algorithm makes computational state in the
target block after the last local queries, it can simply waive the
last local queries since the state must already rest in the target
block before those unnecessary last local iterations. That is, if
$<u|\, \hat{G_2}^{j_{k+1}} \hat{G_1}^{j_{k}} \hat{G_2}^{j_{k-1}}
\cdot \cdot \cdot \hat{G_1}^{j_2} \hat{G_2}^{j_1} \hat{G_1}^{j_0}
\, |s_1>=0$ then we must also have $<u|\, \hat{G_1}^{j_{k}}
\hat{G_2}^{j_{k-1}} \cdot
\cdot \cdot \hat{G_1}^{j_2} \hat{G_2}^{j_1} \hat{G_1}^{j_0} \, |s_1>=0$. \\
2) The first queries could be either local or global iterations.
Sequences starting with the global ($j_0>0$) include:
$\hat{G_1}^{j_0}$ (Global), $\hat{G_1}^{j_2} \hat{G_2}^{j_1}
\hat{G_1}^{j_0}$ (Global-Local-Global), and so on. The simplest
one, with only global queries involved, gives nothing but the
original Grover's full search (\ref{full}), which saves no steps
but precisely locates the target item. The next simplest one,
Global-Local-Global, will be studied in later section. Note that
the up-to-now established optimal partial search GRK algorithm
also falls into this category, with $j_2=1$ and $j_{0},j_1$
specified by
(\ref{korepin}). \\
3) For those starting with local queries ($j_0=0$ and $j_1>0$), we
have: $\hat{G_1}^{j_2} \hat{G_2^{j_1}}$ (Local-Global),
$\hat{G_1}^{j_4} \hat{G_2^{j_3}} \hat{G_1}^{j_2} \hat{G_2^{j_1}}$
(Local-Global-Local-Global), and so on. We will later fully
consider the Local-Global algorithm and also discuss one
particular case of Local-Global-Local-Global which has $j_4$ to be
one, namely with only one global query applied after the last
local iterations (we call this as Local-Global-Local sequence). \\
4) The argument based on the optimality of Grover's full search
 puts lower limit of the partial search steps $S \ge
\frac{\pi}{4} \sqrt{N} - \frac{\pi}{4} \sqrt{b}$. It was shown in
\cite{jaik} that
% only expect partial search to achieve at most $\sqrt{b}$ speedup (at
%large block limit),
$S=\frac{\pi}{4}\sqrt{N}- R \sqrt{b}$. Here $R$ is a constant
independent of $b$. The optimal scheme we are seeking should
maximize $R$. Also according to this thought, we expect the total
number of global queries should scale asymptotically as
$\frac{\pi}{4}\sqrt{N}-\eta \sqrt{b}$, since in partial search we
should be faster than full search $\frac{\pi}{4}$ but save steps
only of order $\sqrt{b}$. As for total number of local queries it
should scale like $\alpha \sqrt{b}$, since $\pi \sqrt{b}$ times
local iteration will have rotated the state vector a whole lap on
the target block $|t> -- |ntt> $ plane. It follows from these
consideration that $R=\eta-\alpha$. Though the numbers of local
and global iterations $j_i$ are integer numbers and in principle
not continuous variables, in the large block limit we can
reasonably treat them as quasi-continuous and use them as function
arguments.
\\
5) $R$ is a function of block number $K$ or instead the parameter
$\gamma=\arcsin(\frac{1}{\sqrt{K}})$ with $0< \gamma \le \pi/4$.
Another limit we will consider is the large $K$ limit, $K>>1$ or
equivalently $\gamma \to 0$. In this limit the mathematical
formulae will be significantly simplified, which will be very
helpful to analytical efforts as will be seen in later sections.
The cases with not so large $K$ could be complemented by direct
numerical verification.

\section{Local-Global sequence of searches }
The Local-Global sequence of searches is the simplest partial
search scheme. For such a sequence $\hat{G}(j_2,j_1)$ with first
$j_1$ local iterations applied and then $j_2$ global, we have the
final computational state to be:
\begin{equation}
\hat{G_1}^{j_2} \hat{G_2}^{j_1} |s_1> = (A,B,C)^T
\end{equation}
Here A,B,C are the coefficients of components $|t>,|ntt>,|u>$
respectively (see (\ref{vectorform})), which are given by
\begin{eqnarray}
 A=&& \sin\gamma \cos(2j_2\theta_1)\sin(2j_1\theta_2) + \sin^2\gamma \sin(2j_2\theta_1)\cos(2j_1\theta_2) + \cos^2\gamma \sin(2j_2\theta_1) \nonumber \\
 B=&& -\sin^2\gamma \sin(2j_2\theta_1)\sin(2j_1\theta_2) + (-)^{j_2} \cos^2\gamma \sin\gamma \cos(2j_1\theta_2) \nonumber \\
   && + \sin^3 \gamma \cos(2j_2\theta_1)\cos(2j_1\theta_2) + \sin\gamma \cos^2\gamma(\cos(2j_2\theta_1-(-)^{j_2})) \nonumber \\
 C=&& -\sin\gamma\cos\gamma\sin(2j_2\theta_1)\sin(2j_1\theta_2) +
 \sin^2\gamma \cos\gamma (\cos(2j_2\theta_1)-(-)^{j_2})\cos(2j_1\theta_2) \nonumber \\
 && + (-)^{j_2} \sin^2\gamma\cos\gamma + \cos^3 \cos(2j_2\theta_1)
\end{eqnarray}
Then to accomplish the partial search we should have the
constraint equation
\begin{equation} <u|\hat{G_1}^{j_2}
\hat{G_2}^{j_1} |s_1> = C = 0
\end{equation}
which means that the amplitude of every item in non-target block
vanish (\ref{constraint}). Since there are only one set of local
and one set of global iterations, we apply the scaling as
discussed before and introduce
\begin{equation}
j_1=\alpha \sqrt{b} \,\,\, , \,\,\, j_2=\frac{\pi}{4}\sqrt{N}-\eta
\sqrt{b}
\end{equation}
We have the limitation $0 \le \alpha < \pi$ and $\eta <
\frac{\pi}{4} \sqrt{K}$. Now we can rewritten the constraint
equation as following:
\begin{equation}
\cos\gamma \cdot {\bigg ( }
[\sin^2\gamma\cos(2\alpha)+\cos^2\gamma]
\sin(\frac{2\eta}{\sqrt{K}})
-\sin\gamma\sin(2\alpha)\cos(\frac{2\eta}{\sqrt{K}}) +(-)^{j_2}
\sin^2\gamma[1-\cos(2\alpha)] { \bigg) } \, = \, 0
\end{equation}
Note that $K\ge 2$ so $\cos\gamma>0$, hence it drops out in the
above equation and we have
\begin{equation}\label{lg_cons}
F(\alpha,\eta)=[\sin^2\gamma\cos(2\alpha)+\cos^2\gamma]
\sin(\frac{2\eta}{\sqrt{K}})
-\sin\gamma\sin(2\alpha)\cos(\frac{2\eta}{\sqrt{K}}) +(-)^{j_2}
\sin^2\gamma[1-\cos(2\alpha)]=0
\end{equation}

Remember we are interested in the fastest algorithm (using the
fewest queries), especially those using less steps than Grover's
full search. Now we have the total number of queries to be
$S=\frac{\pi}{4}\sqrt{N}-(\eta-\alpha)\sqrt{b}$, thus to minimize
$S$ we should maximize $R=\eta-\alpha$ under the constraint
(\ref{lg_cons}). The point here is that from the constraint we can
consider $\eta$ as a function of $\alpha$, thus $R$ is also a
function of $\alpha$, to which it should be optimized. We then
have $\frac{d R}{d\alpha}=-\frac{d\eta}{d\alpha}+1=0$, which leads
to
\begin{equation}
\sin(2\gamma) [ \cos^2\gamma
(1-\cos(2\alpha)\cos(\frac{2\eta}{\sqrt{K}}))+(-)^{j_2} \sin\gamma
\sin(2\alpha)]=0
\end{equation}
Again since $0<2\gamma\le \pi/2$, we have
\begin{equation}
\sin(2\gamma)>0
\end{equation}
so we simplify the above equation as
\begin{equation}\label{lg_dif}
H(\alpha,\eta)=2\sin \alpha [\cos^2\gamma \sin \alpha
\cos(\frac{2\eta}{\sqrt{K}})+(-)^{j_2} \sin \gamma \cos\alpha]=0
\end{equation}

There is an apparent solution for (\ref{lg_cons}) and
(\ref{lg_dif}), namely
$\sin\alpha=\sin(\frac{2\eta}{\sqrt{K}})=0$. This gives $\alpha=0$
and $\eta=l\cdot \pi \cdot \sqrt{K}, l=0,-1,-2,\cdot\cdot\cdot$,
but to maximize $R=\eta-\alpha$ we should have $\alpha=\eta=0$,
which again recovers the Grover's full search solution and is a
trivial one for partial search. In the following we only consider
nontrivial solutions with $\sin\alpha \ne 0$.

Let's first look at the large $K$ or small $\gamma$ limit. In
leading order, the two equations (\ref{lg_cons}) and
(\ref{lg_dif}) are reduced to be
\begin{equation}\label{lg_limit1}
F(\alpha,\eta)=\sin(\frac{2\eta}{\sqrt{K}})-\gamma
\sin(2\alpha)\cos(\frac{2\eta}{\sqrt{K}})+\gamma^2 (-1)^{j_2} 2
 \sin^2 \alpha=0
\end{equation}
\begin{equation}\label{lg_limit2}
H(\alpha,\eta)=\sin\alpha[\sin\alpha
\cos(\frac{2\eta}{\sqrt{K}})+(-1)^{j_2}\gamma \cos\alpha]=0
\end{equation}
To satisfy (\ref{lg_limit1}), the first term
$\sin(\frac{2\eta}{\sqrt{k}})$ must be at least as small as $\sim
\gamma$, so in leading order we must have
$\cos(\frac{2\eta}{\sqrt{k}}) \sim 1$. This, combined with
(\ref{lg_limit2}), requires also $\sin\alpha$ must be at least as
small as $\sim \gamma$ (we don't consider $\sin\alpha=0$ as
mentioned before), which means $\cos\alpha \sim 1$. So finally we
reduce (\ref{lg_limit2}) in leading order to be
\begin{equation}
\sin\alpha + (-1)^{j_2} \gamma =0
\end{equation}
Remember we have $0\le \alpha < \pi$ thus $\sin\alpha \ge 0$, so
the above equation has solution ONLY for ODD values of $j_2$. For
odd $j_2$ the solution is $\sin\alpha=\gamma$, this yields two
solutions $\alpha=\gamma$ and $\alpha=\pi-\gamma$. By substituting
$\sin\alpha=\gamma$ back into (\ref{lg_limit1}) we get in leading
order $\sin{\frac{2\eta}{\sqrt{K}}}=2\gamma^2$, which means
$\frac{2\eta}{\sqrt{K}}=2\gamma^2+2l\pi$ or
$\frac{2\eta}{\sqrt{K}}=-2\gamma^2+(2l-1)\pi$,
$l=0,-1,-2,\cdot\cdot\cdot$. But again remember our goal is to
maximize $R=\eta-\alpha$, so we adopt the optimal Local-Global
solution $\alpha=\gamma$ and $\eta=\gamma$ in large K limit. This,
however, seems giving no speedup compared with full search since
$R=\eta-\alpha \sim 0$. To clarify this, we need go to higher
order of equations (\ref{lg_limit1}) and (\ref{lg_limit2}). By
properly including corrections up to $\sim \gamma^3$ we can find
the solution to be
\begin{equation}
\alpha=\gamma+\frac{1}{2}\gamma^3+o(\gamma^4) \,\,\, , \,\,\,
\eta=\gamma+\frac{2}{3}\gamma^3+o(\gamma^4)
\end{equation}
Now we see in large $K$ limit, the Local-Global sequences of
search do achieve speedup
\begin{equation}
R=\frac{1}{6} \gamma^3 + o(\gamma^4)
\end{equation}
which is very small but still nonzero.

\begin{figure}
\hspace{+0cm}
\centerline{\epsfxsize=13cm\epsffile{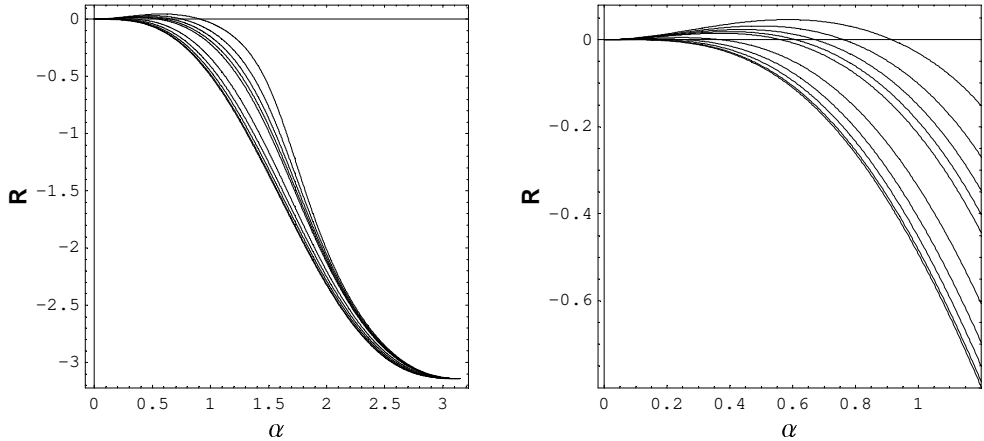}}
\hspace{+0.5cm}
 \caption{\label{fig_LGOdd}
Dependence of $R$ on $\alpha$ with odd $j_2$ in Local-Global
sequence, the curves from top to bottom are for
$K=5,6,7,8,9,18,36,72,144,200$ respectively. For each $\alpha$ the
value of $\eta$ is solved from (\ref{lg_cons}). Right panel is the
amplification of the area near origin in Left panel.}
 \end{figure}

\begin{figure}
\hspace{+0cm}
\centerline{\epsfxsize=8cm\epsffile{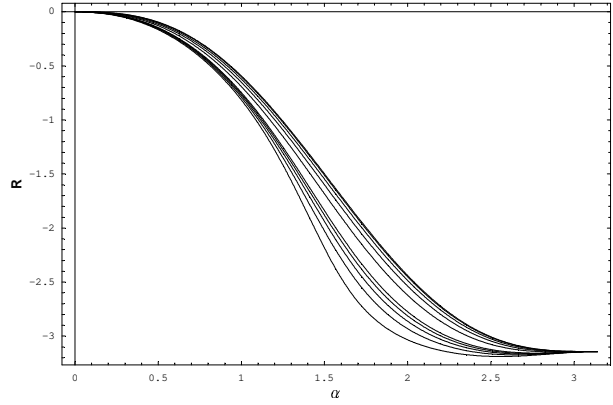}}
\hspace{+0.5cm}
 \caption{\label{fig_LGEven}
Dependence of $R$ on $\alpha$ with even $j_2$ in Local-Global
sequence, the curves from top to bottom are for
$K=5,6,7,8,9,18,36,72,144,200$ respectively. For each $\alpha$ the
value of $\eta$ is solved from (\ref{lg_cons}).}
 \end{figure}

Let's now turn back to finite values of $K$. From (\ref{lg_dif})
we have
\begin{equation} \label{lg_eta}
\cos(\frac{2\eta}{\sqrt{K}})=\frac{(-)^{j_2+1}\sin\gamma\cos\alpha}{\cos^2\gamma\sin\alpha}
\end{equation}
Combine (\ref{lg_eta}) and (\ref{lg_cons}) together we get
\begin{equation}\label{lg_eta2}
\cos(\frac{2\eta}{\sqrt{K}})=\frac{\sin\gamma\cos\alpha}{\cos^2\gamma\sin\alpha}
\,\,\,  , \,\,\, \sin(\frac{2\eta}{\sqrt{K}})=\frac{2\sin^2\gamma
(\cos^2\gamma \sin^2\alpha+\cos^2\alpha)}{\cos^2\gamma
[\sin^2\gamma\cos(2\alpha)+\cos^2\gamma]}
\end{equation}
By requiring
$\cos^2(\frac{2\eta}{\sqrt{K}})+\sin^2(\frac{2\eta}{\sqrt{K}})=1$
we obtain the equation determining the value of $\alpha$ and from
$\alpha$ we can obtain $\eta$. Generally it is hard to analyze the
problem analytically, so we proceed to numerical solutions , see
the figures Fig.\ref{fig_LGOdd},\ref{fig_LGEven}. In these figures
we plot $R$ as a function of $\alpha$ with different $K$ for both
even and odd values of $j_2$, at each point $\eta$ is determined
from the constraint (\ref{lg_cons}). For the odd $j_2$ case there
is always a positive maximum which is faster than full search, and
as $K$ approaches very large values, the maximum of the $\epsilon$
v.s. $\alpha$ curve moves gradually toward the origin which stands
for the full search solution. But for even $j_2$, $R$ is always
negative, hence slower than full search.

An important comparison is to be made between this Local-Global
algorithm and the established GRK one. Our numeric results show
that the odd $j_2$ Local-Global searches can get $R \approx 0.3$
with $K=2,3$ and get $0< R <0.09 $ for $K \ge 4$. For GRK,
however, it can achieve $R > 0.32$ for all $K\ge 2$. So the GRK is
much faster than the present Local-Global searches.

To sum up with the Local-Global sequence of searches, we find it
can be faster than full search with $R \sqrt{b}$ speedup for block
number $K$ up to several tens, but it is always slower than the
GRK optimized partial search.

\section{Global-Local-Global sequence of searches}
In this section we shall consider another algorithm for partial
search, the Global-Local-Global sequence of searches, which starts
with (\ref{ave}), first applies $j_0$ global iterations
(\ref{iter}), then $j_1$ local iterations (\ref{liter}) and
finally $j_2$ global iterations:
\begin{equation}
\hat{G_1}^{j_2} \hat{G_2}^{j_1} \hat{G_1}^{j_0} |s_1\rangle
.\label{glg}
\end{equation}
Again amplitudes of all items in non-target blocks should vanish
at the end of algorithm:
\begin{equation}
\langle u| \hat{G_1}^{j_2} \hat{G_2}^{j_1} \hat{G_1}^{j_0}
|s_1\rangle =0.\label{cancelation}
\end{equation}

Let us present this equation in an explicit form. After first set
of global iterations the state of the database will be:
\begin{equation}
\hat{G_1}^{j_0}|s_1\rangle=\frac{\sin
\gamma}{\sqrt{b}}\left(\begin{array}{c}
\cos (2 j_0 \theta_1)\\
-\sin (2 j_0 \theta_1)\sin \gamma \\
-\sin (2 j_0 \theta_1) \cos \gamma
\end{array} \right)  + \left(\begin{array}{c}
\sin (2 j_0 \theta_1)\\
\sin \gamma \cos (2 j_0 \theta_1)\\
\cos \gamma \cos (2 j_0 \theta_1)
\end{array} \right).
\end{equation}
After local iterations the state of the database is:
\begin{equation}
\hat{G_2}^{j_1} \hat{G_1}^{j_0}|s_1\rangle= \left(\begin{array}{c}
\sin (2 j_0 \theta_1)\cos (2 j_1\theta_2) +\sin \gamma \cos (2 j_0
\theta_1)
\sin (2j_1\theta_2)  \\
-\sin (2 j_0 \theta_1)\sin (2j_1\theta_2)+\sin \gamma \cos (2 j_0 \theta_1)\cos (2j_1\theta_2)\\
\cos \gamma \cos (2 j_0 \theta_1)
\end{array} \right).
\end{equation}
Here we neglected $\sim 1/\sqrt{b}$ terms (namely taking large
block limit). After next set of global searches we have to
calculate only third component(coefficient of $|u\rangle$) of the
vector:
\begin{eqnarray} \label{GLG_cons}
0&=&\langle u|\hat{G_1}^{j_2} \hat{G_2}^{j_1} \hat{G_1}^{j_0} |s_1\rangle   \label{glgmain}\\
&=&\cos (2j_1\theta_2) \left\{ -\cos \gamma \sin
(2j_2\theta_1)\sin (2j_0\theta_1)+ \sin^2\gamma \cos
(2j_0\theta_1)\cos \gamma
\right[(-1)^{j_2+1}+\cos (2j_2\theta_1) \left] \right\} + \nonumber \\
&&\sin (2j_1\theta_2)\left\{-\sin \gamma \cos \gamma \sin
(2j_2\theta_1)\cos (2j_0\theta_1)- \sin \gamma \cos \gamma \sin
(2j_0\theta_1)
 \right[(-1)^{j_2+1}+\cos (2j_2\theta_1) \left] \right\} + \nonumber \\
&& + \cos \gamma  \cos (2j_0\theta_1) [(-1)^{j_2} \sin^2 \gamma +
\cos^2 \gamma \cos (2j_2\theta_1) ] \nonumber
\end{eqnarray}
This constraint equation guarantees the amplitudes of all items in
all non-target blocks vanish to successfully complete the partial
search.

First let us check the case $j_1=0$, no local searches. In this
case the constraint equation can be reduced to $\cos [ 2\theta_1
(j_0+j_2) ] = 0$. This is just the full search: we use
$j_0+j_2=\frac{\pi}{4}\sqrt{N} $ global iterations to find the
target item.

Now let us consider more general and complicated case. We expect
the following scaling, namely the global iterations
$j_0=\frac{\pi}{4}\sqrt{N}-\eta \sqrt{b}$, $j_2=\beta \sqrt{b}$
and total local iterations $j_1=\alpha \sqrt{b}$ with $\eta \le
\frac{\pi}{4} \sqrt{K} $, $0\le \beta < \frac{\pi}{4}\sqrt{K}$ and
$0\le \alpha < \pi$. It should be remembered that our purpose is
to minimize $j_0+j_1+j_2=\frac{\pi}{4} \sqrt{N} - R \sqrt{b} ,\,\,
 R = \eta-\beta-\alpha$, or equivalently maximize $R$. Our
strategy is similar to that used for analyzing Local-Global
sequence: study the large $K$ limit analytically while deal with
finite $K$ case numerically.

In the large $K$ or small $\gamma$ limit, we can take the leading
order of (\ref{GLG_cons}) and simplify our constraint to a much
simpler form :
\begin{equation} \label{GLG_limit}
\eta=\beta \cos(2\alpha)+\frac{1-(-)^{j_3}}{2} \sin(2\alpha)
\end{equation}
From this equation, there are two possibilities: \\
1) $j_2$ is even, thus $\eta=\beta \cos(2\alpha)$. In this case
the problem simplifies into maximizing $R=\beta (\cos(2\alpha)-1)
-\alpha$ with $0\le \beta < \frac{\pi}{4}\sqrt{K}$ and $0\le
\alpha < \pi$. Note that $\cos(2\alpha)-1 \le 0$ so the maximum of
R must occur at $\eta=\beta=\alpha=0$, which is again the trivial
full search solution (\ref{full}).
\\
2) $j_2$ is odd, thus $\eta=\beta \cos(2\alpha)+\sin(2\alpha)$. We
then have to maximize $R=\beta
(\cos(2\alpha)-1)+\sin(2\alpha)-\alpha$ with $0\le \beta <
\frac{\pi}{4}\sqrt{K}$ and $0\le \alpha < \pi$. The solution is
\begin{equation}
\eta=\frac{\sqrt{3}}{2} \,\, , \,\,  \beta=0 \, \, , \, \,
\alpha=\pi/6 ,
\end{equation}
which has achieved $R \sqrt{b} \approx 0.3424 \sqrt{b}$ speedup
with respect to full search. This nontrivial optimal solution is
exactly the GRK algorithm, with vanishingly small odd $j_2$ namely
$j_2=1$.

Now let us discuss  finite values of $K$.
We calculated dependence of  $R$ on $\beta$ 
% We plot $R$ as a function of $\beta$,
 by determining $\alpha$ from constraint equations and
optimizing $\eta$ numerically at each value of $\beta$.
We done this  numerically   for up to 200 blocks.
The dependence of $R$ on  $\beta$ is monotonous.
Corresponding figures took too much memory, so we withdraw them from the paper.
% It can be seen from the figures Fig.\ref{fig_GLGOdd}\ref{fig_GLGEven} that
For even $j_2$ searches we consider  are slower than full Grover
search, while odd $j_2$ these searches are faster than full search. The
optimal searches always occur with odd and vanishing $j_2$ (the
GRK case), being about $0.34\sqrt{b}$ faster than full search. The
large $K$ case also confirm our analysis above. Our results here
also confirm that the  optimum partial search is GRK \cite{kor},
it has the form $\hat{G_1} \hat{G_2}^{j_1}
\hat{G_1}^{j_0}$.

To conclude, we have found that Global-Local-Global sequence of
searches can be much faster than full search (\ref{full}) with odd
times global iterations applied in the end, and the optimal search
is determined to be the GRK algorithm for all values of $K$.

%\begin{figure}
%\hspace{+0cm}
%\centerline{\epsfxsize=13cm\epsffile{fig_glg_odd.eps}}
%\hspace{+0.5cm}
% \caption{\label{fig_GLGOdd}
%Dependence of $R$ on $\beta$ with odd $j_2$ in Global-Local-Global
%sequence, the curves from top to bottom are for
%$K=5,6,7,8,9,18,36,72,144,200$ respectively. For each $\beta$ the
%value of $\alpha$ is solved from (\ref{GLG_cons}) with $\eta$
%optimized numerically. Right panel is the amplification of the
%area near origin in Left panel.}
% \end{figure}

%\begin{figure}
% \centerline{\epsfxsize=8cm\epsffile{fig_glg_even.eps}}
%\hspace{+0.5cm}
% \caption{\label{fig_GLGEven}
%Dependence of $R$ on $\beta$ with even $j_2$ in
%Global-Local-Global sequence, the curves from top to bottom are
%for $K=5,6,7,8,9,18,36,72,144,200$ respectively. For each $\beta$
%the value of $\alpha$ is solved from (\ref{GLG_cons}) with $\eta$
%optimized numerically.}
% \end{figure}

\section{Local-Global-Local sequence of searches}

In this section we shall discuss another algorithm of partial
search. It belongs to the category of Local-Global-Local-Global
sequence but with only one global query applied in the end, which
we call as Local-Global-Local sequence. We start with (\ref{ave})
first apply $j_1$ local iterations, then $j_2$ global iterations
(\ref{liter}) and then $j_3$ local iterations, and eventually a
single global query:
\begin{equation}
\hat{G_1} \hat{G_2}^{j_3}  \hat{G_1}^{j_2} \hat{G_2}^{j_1}
|s_1\rangle .\label{lgl}
\end{equation}
One particular advantage of this type of sequence is that when
$j_1$ tends to zero our Local-Global-Local will degenerate to the
GRK type algorithm. Amplitudes of all items in non-target blocks
should vanish at the end:
\begin{equation}
\langle u| \hat{G_1} \hat{G_2}^{j_3}  \hat{G_1}^{j_2}
\hat{G_2}^{j_1} |s_1\rangle =0.\label{lglcons}
\end{equation}
As before, we introduce the scaling of iteration numbers to be
$j_1=\alpha \sqrt{b}$, $j_3=\delta \sqrt{b}$, and
$j_2=\frac{\pi}{4}\sqrt{N}-\eta\sqrt{b}$. The total number of
queries will be $S=\frac{\pi}{4}\sqrt{N}-R \sqrt{b}$ with
$R=\eta-\alpha-\delta$ which we want minimize under constraint
(\ref{lglcons}). The explicit form of the above constraint
equation is as follows:
\begin{eqnarray}\label{LGL_cons}
0&=&X \cdot \sin(2\alpha) + Y \cdot {\big [ } \cos(2\alpha)-1 {\big ]} + Z \\
  && X=\sin\gamma \sin(2\gamma) \sin(2\delta) \sin(\frac{2\eta}{\sqrt{K}}) + \sin^2\gamma \sin(2\gamma)
   \cos(2\delta) \cos(\frac{2\eta}{\sqrt{K}}) + \sin\gamma \cos\gamma \cos(2\gamma) \cos(\frac{2\eta}{\sqrt{K}})  \nonumber \\
  && Y= \sin^2\gamma \sin(2\gamma) \sin(2\delta) \cos(\frac{2\eta}{\sqrt{K}}) - \sin^3\gamma \sin(2\gamma)\cos(2\delta)\sin(\frac{2\eta}{\sqrt{K}})
        -\sin^2\gamma \cos\gamma {\bigg [} \cos(2\gamma)\sin(\frac{2\eta}{\sqrt{K}})+(-)^{j_2} {\bigg ]}   \nonumber \\
  && Z= \sin(2\gamma) \sin(2\delta) \cos(\frac{2\eta}{\sqrt{K}}) - \sin\gamma \sin(2\gamma) \cos(2\delta) \sin(\frac{2\eta}{\sqrt{K}})
         -\cos\gamma \cos(2\gamma) \sin(\frac{2\eta}{\sqrt{K}}) \nonumber
\end{eqnarray}
Though the above equation is complicated, it is very easy to solve
numerically. Also taking large $K$ limit can significantly
simplify it. So as we did before, we analytically study the large
$K$ limit, complemented by numerical results from very small $K$
to very large $K$.

By taking large $K$ or small $\gamma$ limit, we obtain from
leading order of (\ref{LGL_cons}) the following:
\begin{equation} \label{LGL_conssim}
\sin(2\alpha)+2\sin(2\delta)-2\eta=0
\end{equation}
We then have $\eta=\frac{1}{2} \sin(2\alpha) + \sin(2\delta)$ and
hence the total number of queries to be
\begin{equation}
R=\frac{1}{2}\sin(2\alpha)-\alpha+\sin(2\delta)-\delta
\end{equation}
By requiring $\frac{\partial R}{\partial \alpha}=\frac{\partial
R}{\partial \delta}=0$ we get the solution maximizing $R$
\begin{equation}
\alpha=0 \,\,\, , \,\,\, \delta=\frac{\pi}{6} \,\,\, , \,\,\,
\eta=\frac{\sqrt{3}}{2}
\end{equation}
This, with zero $j_1$, again recovers the GRK optimized solution.
So in large $K$ limit we see Local-Global-Local sequence is no
faster than GRK algorithm. Different from Local-Global and
Global-Local-Global, here in Local-Global-Local we notice that the
odd $j_2$ and even $j_2$ converge to each other when approaching
the optimal solution and the oscillation terms in (\ref{LGL_cons})
with factor $(-)^{j_2}$ disappear.

\begin{figure}
\hspace{+0cm}
\centerline{\epsfxsize=13cm\epsffile{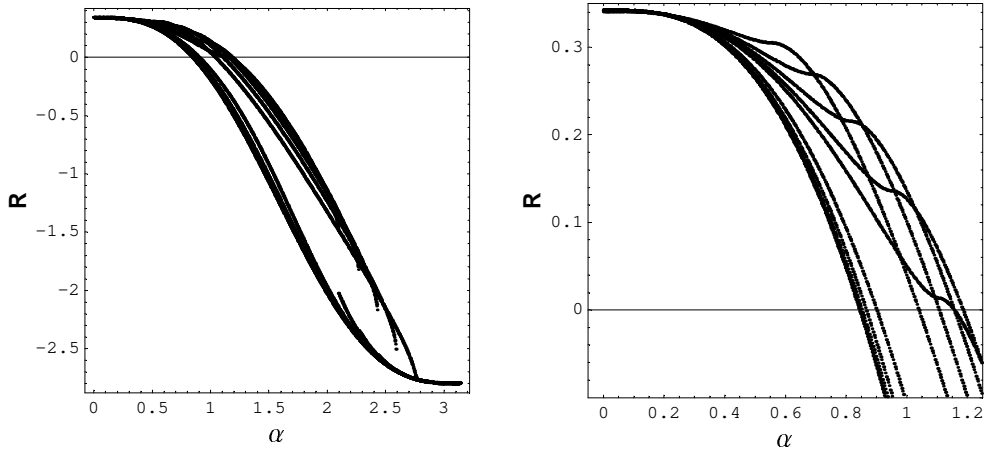}}
\hspace{+0.5cm}
 \caption{\label{fig_LGLOdd}
Dependence of $R$ on $\alpha$ with odd $j_2$ in Local-Global-Local
sequence, the curves from top to bottom are for
$K=5,6,7,8,9,18,36,72,144,200$ respectively. For each $\alpha$ the
value of $\eta$ is solved from (\ref{LGL_cons}) with $\delta$
optimized numerically. Right panel is the amplification of the
area near origin in Left panel.}
 \end{figure}

\begin{figure}
 \centerline{\epsfxsize=13cm\epsffile{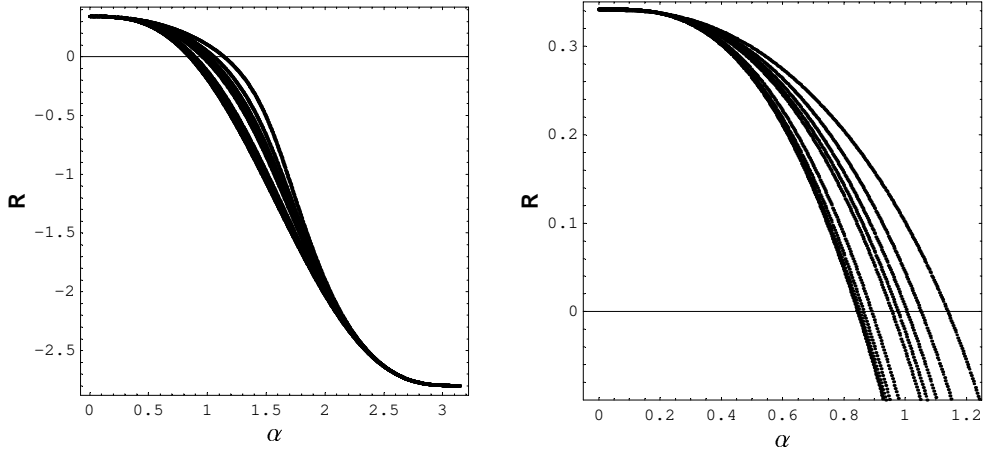}}
\hspace{+0.5cm}
 \caption{\label{fig_LGLEven}
Dependence of $R$ on $\alpha$ with even $j_2$ in
Local-Global-Local sequence, the curves from top to bottom are for
$K=5,6,7,8,9,18,36,72,144,200$ respectively. For each $\alpha$ the
value of $\eta$ is solved from (\ref{LGL_cons}) with $\delta$
optimized numerically.}
 \end{figure}

For general values of $K$, we conduct the numerical method and
show the results in figures Fig.\ref{fig_LGLOdd}\ref{fig_LGLEven},
which plot $R$ as a function of $\alpha$ with $\eta$ solved from
(\ref{LGL_cons}) and $\delta$ optimized numerically. As can be
seen, the optimal solutions always occur with $\alpha=0$ which
goes back to GRK case. Also we note that even and odd $j_2$ give
same results around optimal point.

So in this section we have established that the Local-Global-Local
sequence of searches can be much faster than full search, but is
no faster than GRK algorithm. In the appendix an alternative
approach for Local-Global-Local sequence based on a conjecture
about cancellation of oscillation terms in (\ref{LGL_cons}) is
briefly described, which though is not directly relevant here but
arrives at similar result and may shed light for future
exploration of even more complicated sequences.

\section{Summary}
We considered different partial search algorithms, which consists of a 
sequence of local and global searches.
We introduced  a general framework for studying partial
quantum search algorithms and  classified various possible sequences.
Particularly, we studied  the Local-Global,
Global-Local-Global as well as Local-Global-Local sequences of
searches by combining numerical study for wide range values of $K$
and analytical results for large $K$ limit. All these algorithms
 achieve $\sqrt{b}$ speedup compared to the  Grover's full quantum
search. GRK algorithm \cite{kor} is the fastest among 
partial search algorithms, which we considered.

\section*{Acknowledgements}

 The paper was supported by NSF Grant DMS-0503712.

\appendix
\section*{Appendix: remarks on
Local-Global-Local sequence}

Let us make two remarks:

1) Let's start from the explicit form of constraint equation
(\ref{LGL_cons}). We  minimizing $j_1+j_2+j_3$. It is interesting
to note  that at the minimum the oscillation terms cancel. The
coefficient at $(-1)^{j_2}$ vanishes because of GRK equation:
\begin{equation}
\cos \gamma \sin 2 \gamma  \cos (2\theta_2j_3)= \sin  \gamma \cos
2 \gamma \label{uda}
\end{equation}
This is exactly the first equation of (\ref{korepin}). We also can
write it in the form:
\begin{equation}
\cos (2j_3\theta_2)= \frac{1-\tan^2 \gamma}{2} , \qquad \sin
(2j_3\theta_2)=\frac{\sqrt{3-4\sin^2 \gamma}}{2\cos^2 \gamma}
\label{uda}
\end{equation}

2)  Minimum number of iterations for Local-Global-Local sequences
corresponds to $j_1=0$ case and the algorithm is reduced back to
GRK version of partial search, see small $j_1$ increase $j_2+j_1$
very little since $d(j_2+j_1)/dj_1=0$ and $d^2(j_2+j_1)/dj_1^2=0$
at  $j_1=0$.

\end{document}